\documentclass[aps,prb,superscriptaddress,amsfonts,amsmath,amssymb,twocolumn]{revtex4}
\usepackage{bm}
\usepackage{graphicx}
\usepackage{amsmath}
\usepackage{amstext}
\usepackage{amssymb}
\usepackage{amsfonts}
\usepackage{amsbsy}
\usepackage{verbatim}
\usepackage{feyn}

\def\br{\boldsymbol{r}}

\def\rSU{{\rm SU}}

\def\rSU2{{\rm SU}(2)}

\begin{document}

\title{Erratum: algebraic spin liquid as the mother of many competing orders}

\author{Michael Hermele}
\affiliation{Department of Physics, University of Colorado, Boulder, Colorado 80309, USA}
\author{T. Senthil}
\affiliation{Department of Physics, Massachusetts Institute of Technology, Cambridge, Massachusetts 02139, USA}
\author{Matthew P. A. Fisher}
\affiliation{Microsoft Research, Station Q, University of California, Santa Barbara, California 93106, USA}
\affiliation{Physics Department, University of California, Santa Barbara, California 93106, USA}

\date{\today}
\begin{abstract}
We correct an error in our paper Phys. Rev. B 72, 104404 (2005) [cond-mat/0502215].  We show that a particular fermion bilinear is not related to the other ``competing orders'' of the algebraic spin liquid, and does not possess their slowly decaying correlations.  For the square lattice staggered flux spin liquid (equivalently, $d$-wave RVB state), this observable corresponds to the uniform spin chirality.
\end{abstract}
\maketitle

In a recent paper,\cite{hermele05} we studied the staggered flux algebraic spin liquid state of the square lattice $S = 1/2$ Heisenberg antiferromagnet.  We emphasized the presence of an ${\rm SU}(4)$ emergent symmetry present at low energy in this state, and showed that it leads to a striking unification of several seemingly unrelated ``competing orders,'' including order parameters for the Neel and valence bond solid states.  In the field theory of the algebraic spin liquid, which consists of $N_f = 4$ two-component Dirac fermions coupled to a ${\rm U}(1)$ gauge field, these competing orders arise as a set of fermion bilinears transforming in the adjoint representation of ${\rm SU}(4)$: $N^a = \Psi^\dagger \tau^3 T^a \Psi^{\vphantom\dagger}$.  These observables exhibit slow power law correlations, characterized by the power law decay $\langle N^a(\br) N^b(0) \rangle \sim \delta_{a b} / |\br|^{2 \Delta_N}$, where one expects $\Delta_N$ is smaller than its large-$N_f$ (or mean field) value of 2.  This expectation is based on the result of Rantner and Wen that $\Delta_N = 2 - 64/ ( 3 \pi^2 N_f ) + {\cal O}(1/N_f^2)$,\cite{rantner02} as well as the  physical picture that the ${\rm U}(1)$ gauge force tends to bind the oppositely charged $\Psi$ and $\Psi^\dagger$ fermions together.  This makes the mode created by $N^a$ propagate more like a single particle-like boson, as opposed to a composite of two free fermions, thus reducing $\Delta_N$.  This tendency increases as $N_f$ is decreased, which reduces the screening of the gauge field and leads to a stronger gauge interaction between fermions.

In Ref.~\onlinecite{hermele05},  we claimed that the ${\rm SU}(4)$ singlet 
$M = \Psi^\dagger \tau^3 \Psi^{\vphantom\dagger}$ has the same power law decay as $N^a$, that is $\langle M(\br) M(0) \rangle \sim 1 / |\br|^{2 \Delta_M}$, where $\Delta_M = \Delta_N$ to all orders in the $1/N_f$ expansion.  However, our argument in Appendix D of Ref.~\onlinecite{hermele05} missed an important class of diagrams, and as a result this statement is not correct.  In fact, $\Delta_M$ and $\Delta_N$ differ at order $1/N_f$.  Here, we outline the calculation of $\Delta_M$ to this order.  The result is $\Delta_M = 2 + 128 /(3 \pi^2 N_f)  + {\cal O}(1/N_f^2)$.  Next, we discuss this result in the context of the physical picture of gauge binding given above.

We calculate $\Delta_M$ by adding $m M$ as a perturbation to the action, and examining its leading order contribution to the fermion Green's function; this is along the same lines as the calculation of the scaling dimension of velocity anisotropy in Appendix C of Ref.~\onlinecite{hermele05}.  The vertex corresponding to insertion of $M$ is denoted by 
\begin{equation}
\feyn{ff}\annotate{-1.5}{-0.1}{\blacktriangleright}\annotate{-0.5}{-0.1}{\blacktriangleright}\annotate{-1.0}{-0.1}{\otimes} = i m \text{.}
\end{equation}
We consider the term in the fermion self-energy of first order in both $1/N_f$ and in $m$, denoted $\Sigma(k) = \sum_{i = 1}^3 \Sigma_i(k)$.  We have
\begin{equation}
\Sigma_1(k) = \feyn{fglf}
\annotate{-1.5}{-0.1}{\blacktriangleright}
\annotate{-0.5}{-0.1}{\blacktriangleright}\annotate{-1.0}{-0.1}{\otimes} \text{.}
\end{equation}
A corresponding diagram is also present for $N^a$, and leads to the result of Ref.~\onlinecite{rantner02}.

In addition, we have
\begin{equation}
\Sigma_2(k) = \feyn{gvffsgv}
\annotate{-0.8}{0.9}{\feyn{flflu}}
\annotate{-0.8}{-0.1}{\blacktriangleright}
\annotate{-0.8}{0.35}{\blacktriangleright}
\annotate{-0.8}{1.4}{\otimes} \text{,}
\end{equation}
and
\begin{equation}
\Sigma_3(k) = \feyn{gvffsgv}
\annotate{-0.8}{0.9}{\feyn{flflu}}
\annotate{-0.8}{-0.1}{\blacktriangleright}
\annotate{-0.8}{0.35}{\blacktriangleleft}
\annotate{-0.8}{1.4}{\otimes} \text{.}
\end{equation}
These diagrams, which were missed by us in Ref.~\onlinecite{hermele05}, only contribute to $\Delta_M$;  the corresponding diagrams for $N^a$ vanish because the fermion loop involves a trace over a single
${\rm SU}(4)$ generator.

These diagrams can be evaluated using dimensional regularization, as in Appendix C of Ref.~\onlinecite{hermele05}.  Keeping only the logarithmically divergent parts, we have
\begin{eqnarray}
\Sigma_1(k) &=& - \frac{24}{\pi^2 N_f} (i m) \operatorname{ln} \Big( \frac{|k|}{\mu} \Big) \\
\Sigma_2(k) &=& \Sigma_3(k) =  \frac{32}{\pi^2 N_f} (i m) \operatorname{ln} \Big( \frac{|k|}{\mu} \Big) \text{.}
\end{eqnarray}
Applying the Callan-Symanzik equation as described in Ref.~\onlinecite{hermele05}, we find
\begin{equation}
\Delta_M = 2 + \frac{128}{3 \pi^2 N_f} + {\cal O}(1/N_f^2) \text{.}
\end{equation}

It is perhaps surprising that $\Delta_M$ has \emph{increased} above its $N_f \to \infty$ value, because this seems to call into question the validity of the physical picture that gauge binding should lower the scaling dimension.  However, there is an important difference between the $N^a$ and $M$ fermion bilinears.  The $M$ fermion bilinear can decay into photons, as represented by the diagram
\begin{equation}
\feyn{flflu}
\annotate{1.5}{-0.5}{\feyn{gg}}
\annotate{1.5}{0.3}{\feyn{gg}}
\annotate{-0.95}{-0.1}{\otimes}
\annotate{0.0}{-0.87}{\blacktriangleright}
\annotate{0.0}{0.7}{\blacktriangleleft}
\annotate{2.4}{-0.1}{\text{.}}
\end{equation}
 This is impossible for $N^a$, because it carries the ${\rm SU}(4)$ flavor quantum number.  Such decay processes will reduce the ability of the mode created by $\Psi^\dagger \tau^3 \Psi^{\vphantom\dagger}$ to propagate as a single particle-like object, and thus compete against the binding due to the gauge force.  Evidently, at least at large-$N_f$ the decay processes win out over the gauge binding, and $\Delta_M$ is increased above the value for noninteracting fermions.  Because no similar decay process is allowed for $N^a$, we still expect $\Delta_N < 2$ for all values of $N_f$.
 
We would like to thank Subir Sachdev for bringing this error to our attention.
 
\bibliography{erratum}
 
\end{document}